\documentclass[10pt,twocolumn]{article}
\usepackage{times}
\pdfoutput=1
\newenvironment{BoldAbstract}{
\begin{quote} \bf}
{\end{quote}}

\usepackage{amsmath}
\usepackage{amssymb}
\usepackage{amsfonts}
\usepackage{graphicx}
\usepackage{epsfig}
\usepackage{color}
\usepackage[style=nature]{biblatex}
\usepackage{scicite}
\usepackage[labelfont=bf]{caption}

\newcommand{\I}[1]{\textrm{Im}\{#1\}}

\newcommand{\etal}{\emph{et al}. }
\def\f12{\frac{1}{2}}

\long\def\symbolfootnote[#1]#2{\begingroup%
 \def\thefootnote{\fnsymbol{footnote}}\footnotetext[#1]{#2}\endgroup}

\def\Eb{{\rm \bf E}}  \def\pb{{\rm \bf p}}
  \def\rb{{\rm \bf r}}
\def\ii{{\rm  i}}   \def\ee{{\rm  e}} \def\dd{{\rm  d}}

\title{Rotational Doppler cooling and heating}

\author{Deng~Pan,$^{1}$\footnotemark$\,\,$ Hongxing~Xu$^{2}$ and F.~Javier~Garc\'{\i}a~de~Abajo$^{1, 3}$\\
\normalsize{$^{1}$ ICFO-Institut de Ciencies Fotoniques, The Barcelona Institute of Science and Technology,}\\
\normalsize{08860 Castelldefels (Barcelona), Spain.} \\
\normalsize{$^{2}$ School of Physics and Technology, Wuhan University, Wuhan 430072, China.}\\
\normalsize{$^{3}$ ICREA-Instituci\'o Catalana de Recerca i Estudis Avan\c{c}ats, Passeig Llu\'{\i}s Companys 23,}\\
\normalsize{08010 Barcelona, Spain.}
}

\date{}

\begin{document}
\twocolumn[
  \begin{@twocolumnfalse}
    \maketitle
    		\begin{BoldAbstract}
Doppler cooling \cite{Wineland1975,Hansch1975,Wineland1978} is a widely used technique to laser cool atoms and nanoparticles exploiting the Doppler shift involved in translational transformations. The rotational Doppler effect \cite{Garetz1981,Korech2013} arising from rotational coordinate transformations should similarly enable optical manipulations of the rotational degrees of freedom in rotating nanosystems. Here, we show that rotational Doppler cooling and heating (RDC and RDH) effects embody rich and unexplored physics, such as a strong dependence on particle morphology. For geometrically confined particles, such as a nanorod that can represent diatomic molecules, RDC and RDH follow similar rules as their translational Doppler counterpart, where cooling and heating are always observed at red- or blue-detuned laser frequencies, respectively. Surprisingly, nanosystems that can be modeled as a solid particle shows a strikingly different response, where RDH appears in a frequency regime close to their resonances, while a detuned frequency produces cooling of rotation. We also predict that the RDH effect can lead to unprecedented spontaneous chiral symmetry breaking, whereby an achiral particle under linearly polarized illumination starts spontaneously rotating, rendering it nontrivial compared to the translational Doppler effect. Our results open up new exciting possibilities to control the rotational motion of molecules and nanoparticles.
             \end{BoldAbstract}
  \end{@twocolumnfalse}
  ]

\symbolfootnote[1]{Electronic address: deng.pan@icfo.eu}

Radiation pressure arises on a surface when it reflects or absorbs light. This phenomena was theoretically predicted by Maxwell and experimentally demonstrated over a century ago \cite{Lebedew1901,Nichols1903,Nichols1903_2}, although the observed radiation pressure was found to be feeble. The advent of lasers gave birth to various techniques to trap and manipulate small particles and macroscopic objects using by optical forces \cite{Ashkin1970,Ashkin1971,Ashkin1986}. Radiation pressure can also be used to reduce Brownian motion of a mirror, provided one supplies an active feedback to preserve the direction of the optical force opposing instant Brownian velocities \cite{Cohadon1999,Arcizet2006}. Also important is the passive technique to control the temperature of atoms and nanostructures, in which a laser of frequency red- or blue-detuned with respect to the intrinsic frequency of atoms \cite{Wineland1978,Chu1985} or optical cavities \cite{Chan2011,Arcizet2006_2} always leads to cooling and heating, respectively. Doppler cooling, which lays the foundation of various passive cooling techniques, was first achieved for atoms\cite{Wineland1978,Chu1985} and recently for molecules \cite{Shuman2010,Hummon2013,Anderegg2018}. Compared with cooling, the passive heating effect is less explored, and can lead to mechanical instabilities of a cavity \cite{Arcizet2006_2}.

Apart from the optical control of translational motions, optical forces can also affect rotational motion of particles with rotational degrees of freedoms, such as molecules or nanoparticles. Circularly polarized light with intrinsic angular momentum can exert a torque to accelerate the rotation of particles, while rotation frequencies as high as GHz have been already achieved for particles up to micron-size in ultra-high vacuum environment \cite{Reimann2018,Ahn2018}. In the context of temperature control by lasers, although rovibronic transitions \cite{Tong2010} are also addressed during laser cooling targeting translational motion, the fundamentals of laser cooling and heating of rotational degrees of freedom have not been clearly revealed.

\begin{figure}[t]
\includegraphics[width=1.0\columnwidth]{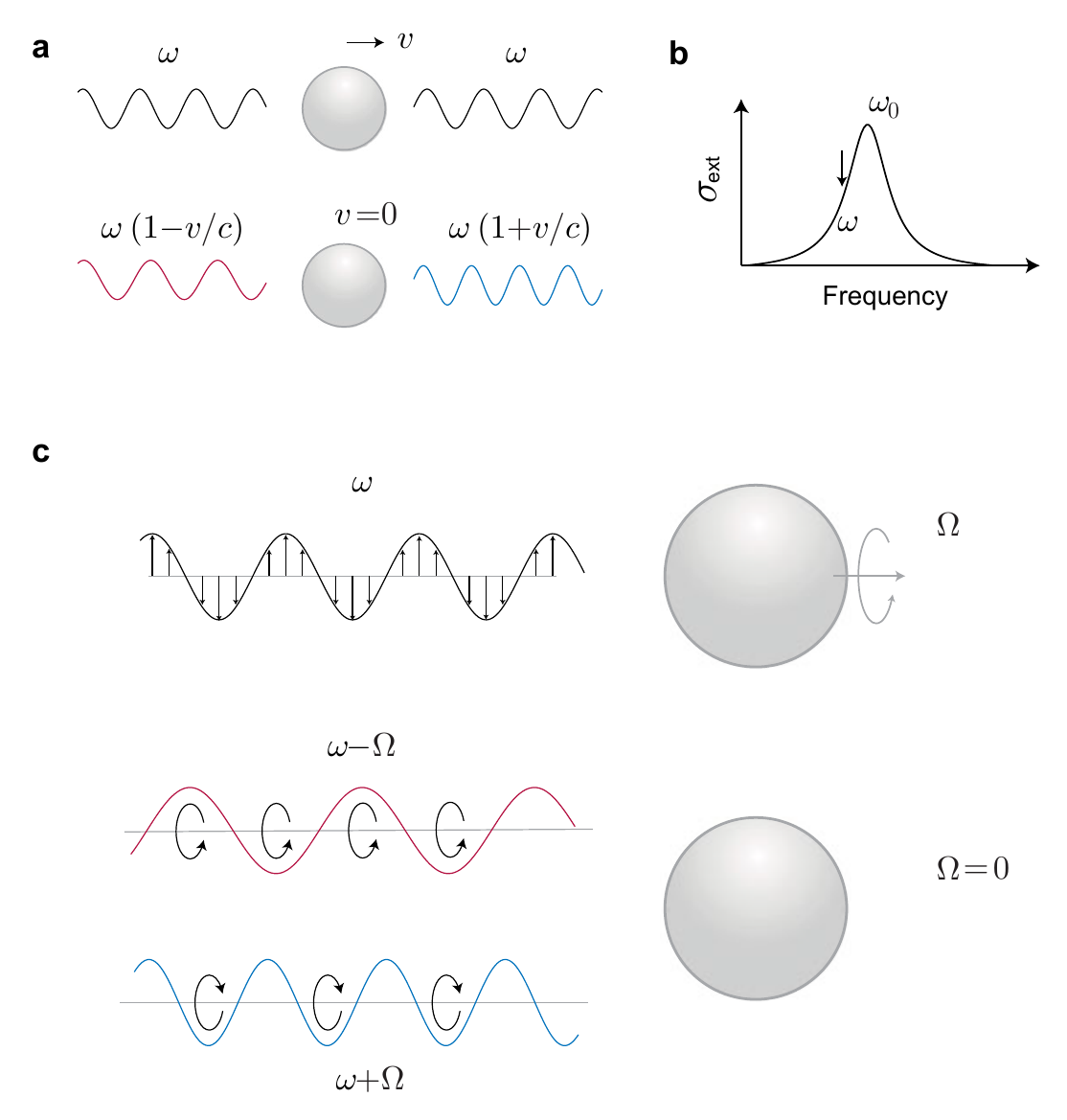}
\caption{\small {\bf Comparison between translational and rotational Doppler cooling and heating.} {\bf a,b,} Illustration of translational Doppler cooling. In the lab frame (upper scheme, {\bf a}), the particle is moving at a velocity $v$ in the presence of two counter-propagating light waves (curves) of frequency $\omega$. In the frame moving with the particle (lower scheme, {\bf a}), the extinction cross section ($\sigma_{\rm ext}$) of these two waves (see {\bf b}), is affected by Doppler shifts to $\omega(1\pm v/c)$, leading to an optical force that pushes the particle to the left, thus causing particle deceleration in the lab frame. {\bf c,} A particle rotating with angular velocity $\Omega$ is illuminated by linearly polarized light of frequency $\omega$(upper scheme). The incident light can be decomposed into RCP and LCP components, which are rotation-Doppler shifted to $\omega\pm\Omega$ in the frame rotating with the particle (lower scheme).
} \label{Fig1}
\end{figure}

Here, we generalize the Doppler cooling and heating effects, usually discussed for translational motion as schematically illustrated in Fig.\ \ref{Fig1}a,b, to the rotational degrees of freedom (see Fig.\ \ref{Fig1}c). In Fig.\ \ref{Fig1}a, we show a particle moving with velocity ${\bf v}$ and illuminated by two counter-propagating light waves of equal intensity and frequency $\omega$ (upper panel, Fig.\ \ref{Fig1}a). In the frame moving with the particle (lower panel, Fig.\ \ref{Fig1}a), the two light waves propagating parallel and anti-parallel relative to $\bf v$ are red- and blue-Doppler-shifted, respectively, to $\omega(1\pm v/c)$. Assuming a red-detuned laser frequency $\omega$ relative to a dominant particle resonance $\omega_0$ (Fig.\ \ref{Fig1}b), compared with the red-shifted laser, the blue-shifted anti-parallel laser is closer to $\omega_0$ and thus undergoes stronger scattering, resulting in deceleration and cooling of the particle. Similarly, heating can be achieved through blue-detuning.

To generalize this scheme to rotational motion, we consider a particle trapped by a linearly polarized beam (upper panel, Fig.\ \ref{Fig1}c). Right and left circularly polarized (RCP and LCP) components of the incident light have then equal intensity, thus resembling the two counter-propagating waves in Fig.\ \ref{Fig1}a. Consequently, they experience different rotational Doppler shifts to $\omega_\pm=\omega\pm\Omega$ when transforming the system to the frame rotating with the particle (lower panel, Fig.\ \ref{Fig1}c). Similar to the translational dynamics, the direction of the torque acting on the particle is then determined by the relative magnitude of the torque exerted by RCP and LCP components. One might naively conclude that the rotational Doppler cooling and heating effects (RDC and RDH) should also appear at red- and blue-detuned $\omega$. However, we show below that this conclusion is only valid for particles of certain geometries. In this respect, rotational transformations are non-inertial, so that the optical response of the rotating particle is more complicated and thus contains a richer physics. More importantly, the trapping of the particle at the antinode of standing waves by a gradient force cannot be extrapolated to the scenario of rotation, which as we explain below can lead to instabilities of the particle at rest ($\Omega=0$) for laser frequencies in the heating regime.

%
To rigorously calculate the torque exerted on rotating particles by linearly polarized light, in what follows, we first find the polarizabilities of particles with different geometries by a simple classical model, which describes the dipolar optical mode as a harmonically oscillating charge (mass $m$, charge $Q$) driven by the electric field of light. Such classical description is capable of adequately addressing transition dipole moments, such as in molecules, and in fact it also satisfies the optical theorem, while a naive application of first-order quantum theory fails to comply with that theorem \cite{Berman2006}. In term of the radial position vector of the charge $\rb$, the classical equation of charge motion becomes
\begin{align}
\ddot{\rb}=-\omega_0^2 r \hat{r}-\gamma (\dot{\rb}-\Omega r \hat{\varphi})+\tau\dddot{\rb}+\frac{Q}{m}\Eb+{\bf F}^{\rm react}      ,\label{motion}
\end{align}
where $\omega_0$ is the intrinsic oscillator resonance frequency, internal dissipation is captured by a phenomenological damping rate $\gamma$ appearing in a term that is proportional to the velocity with respect to the center of mass $\dot{\rb}-\Omega r \hat{\varphi}$, the Abraham-Lorentz force $m\tau\dddot{\rb}$ with $\tau=2Q^2/3mc^3$ introduces minor corrections due to radiation, and ${\bf F}^{\rm react}$ is the force imposed by the boundary defined by the particle geometry.

\begin{figure*}[t]
\includegraphics[width=2.0\columnwidth]{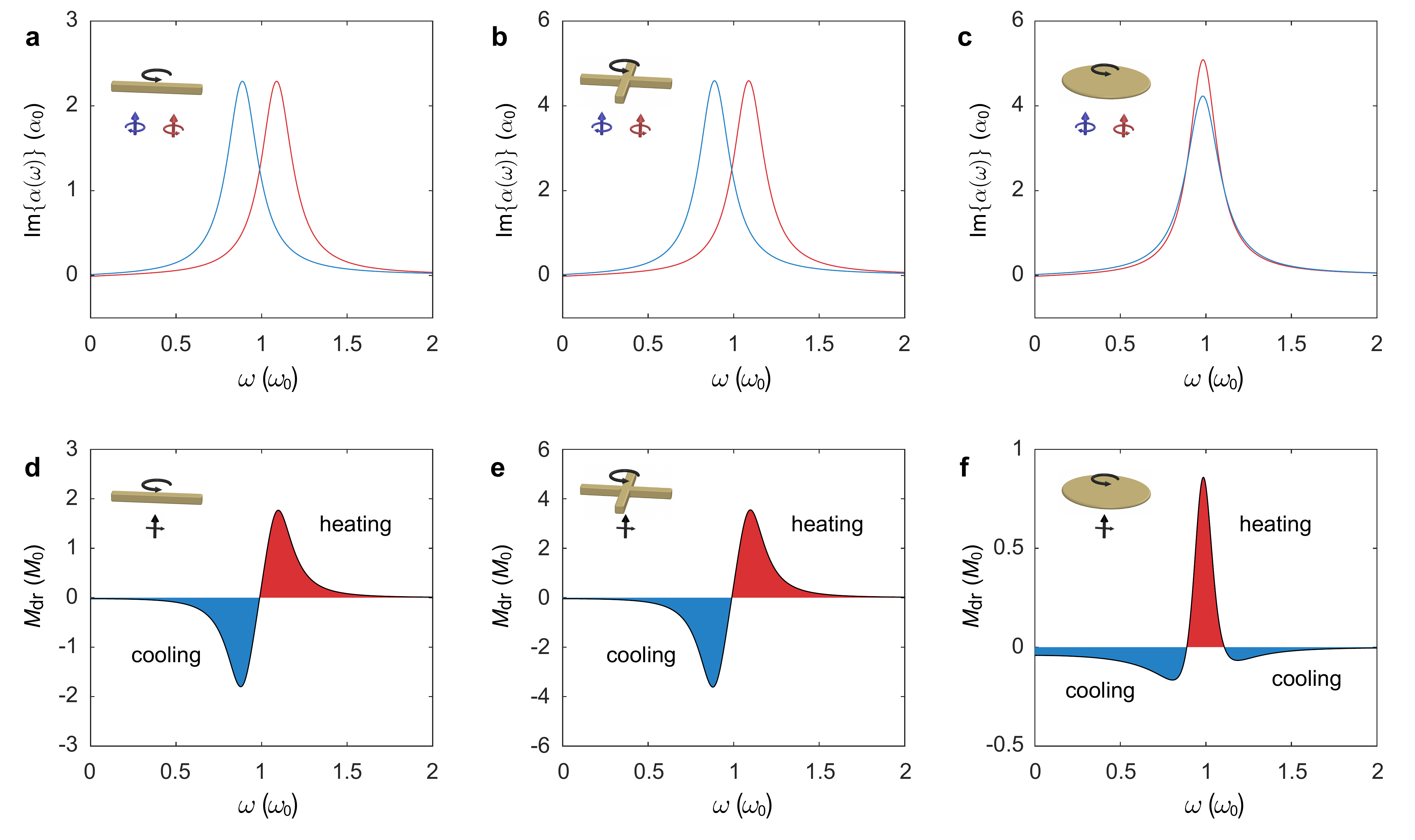}
\caption{\small {\bf Optical torque on rotating particles.} {\bf a,b,c} Optical response of rotating particles of different geometries, namely a nanorod ({\bf a}), a nanocross ({\bf b}) and a nanodisk ({\bf c}), to RCP (red arrows) and LCP (blue arrows) incident light. We plot the imaginary part of the polarizability, which is related to the extinction cross section according to $\sigma_{\rm ext}=4\pi k\I{\alpha}$. A solid particle such as the nanodisk ({\bf c}) exhibits different CD with particles relative to confined electron motions, such as the nanorod ({\bf a}) and nanocross ({\bf b}) {\bf d,e,f} Time-average torque acting on the rotating particles considered in {\bf a}, {\bf b} and {\bf c}. In all cases, particles are rotating with angular velocity, damping rate $\gamma=0.2\omega_0$ and $\tau=0.02 \omega_0^{-1}$. All frequencies are normalized to the particle resonance frequency $\omega_0$, the polarizability is normalized to $\alpha_0=Q^2/m\omega_0^2$, and the torque is normalized to $M_0=\alpha_0 c|E_\pm|^2/2\pi\omega_0$.
} \label{Fig2}
\end{figure*}

In a thin nanorod (Fig.\ \ref{Fig2}a), the oscillating dipole $\pb$ produced by the bounded charge is oriented along the rod axis and rotating with the particle. For RCP ($+$) and LCP ($-$) light with electric field $\Eb_\pm=(\hat{x}\pm \ii\hat{y})E_\pm\ee^{-\ii\omega t}/\sqrt{2}$, we can define an effective polarizability that relates the field component parallel with the nanorod to the dipole moment $p_\pm=2\alpha^\pm_{\rm rod}\Eb_\pm\cdot\hat{\rb}$. More precisely, we find
\begin{align}
\alpha^\pm_{\rm rod}(\omega)=\frac{Q^2/2m}{\omega_0^2-\Omega^2-\omega_\mp^2 - i\Gamma_\mp}   \label{rod}
\end{align}
with
\begin{align}
\Gamma_\pm= \gamma \omega_\pm +\tau\omega_\pm (\omega_\pm^2+3\Omega^2).  \nonumber
\end{align}

For an optically isotropic nanoparticle, the circular polarizability is determined by $\pb_\pm=\alpha^\pm\Eb_\pm$, where $\pb_\pm$ is formed by two orthogonal degenerate dipole moments. One example of optical isotropic particle is illustrated by connecting two nanorod at their centers as shown in Fig.\ \ref{Fig2}b, where we assume a charge $Q$ in each of the two orthogonal directions. Applying Eq.\ (\ref{motion}) to this nanocross, we find
\begin{align}
\alpha^\pm_{\rm cross}(\omega)=\frac{Q^2/m}{\omega_0^2-\Omega^2-\omega_\mp^2 -i(\gamma\omega_\mp+\tau\omega_\mp^3)}      .\label{cross}
\end{align}

For isotropic nanoparticles containing freely moving electron, such as nanodisk shown in Fig.\ \ref{Fig2}c, we need to consider the coupling between the two orthogonal charge oscillators through a Coriolis force, which leads to the polarizability
\begin{align}
\alpha^\pm_{\rm disk}(\omega)= \frac{Q^2/m}{\omega_0^2-2\Omega^2-\omega^2 - i(\gamma\omega_\mp+\tau\omega^3)}              .\label{disk}
\end{align}

The polarizabilities obtained in Eqs.\ (\ref{rod}), (\ref{cross}) and (\ref{disk}) can fully describe the optical response of small rotating particles and the optical torque produced under external illumination. According to the optical theorem, the extinction cross sections of the particles are determined by $\sigma_{\rm ext}^\pm=4\pi k\I{\alpha^\pm}$, where $k=\omega/c$. However, the elastic scattering process, as described by the cross section $\sigma_{\omega}^\pm=8\pi k^4 |\alpha^\pm|^2/3$, maintains the angular momentum of light and thus does not lead to a torque on the particle. In contrast, each absorbed circularly polarized photon directly transfers angular momentum $\hbar$ to the particle, so we are interested in the absorption cross section $\sigma_{\rm abs}^\pm=\sigma_{\rm ext}^\pm-\sigma_{\omega}^\pm-\sigma_{\omega\mp 2\Omega}^\pm$. Here, $\sigma_{\omega\mp 2\Omega}^\pm$ is the cross section of inelastic scattering (the so-called rotational Doppler scattering or rotational Raman scattering), which is only present in an rotating anisotropic particle and accompanied by an exchanging of AM $2\hbar$ for each scattered photon. For a nanorod we can find $\sigma_{\omega\mp 2\Omega}^\pm=8\pi(\omega\mp 2\Omega)^4 |\alpha^\pm|^2/3c^4$. Following these considerations and writing intensity of incident light as $I=c|E_\pm|^2/8\pi$, the total torque exerted on the particle by RCP and LCP components reduces to
\begin{align}
M_\pm=\pm(2\sigma_{\omega\mp 2\Omega}^\pm+\sigma_{\rm abs}^\pm)|E_\pm|^2/2\pi k   .\label{torque}
\end{align}

Figure \ref{Fig2}a,b,c shows imaginary part of the circular polarizabilities $\I{\alpha^\pm}$ for three types of rotating particles according to Eqs.\ (\ref{rod}), (\ref{cross}) and (\ref{disk}). We observe a strong circular dichroism (CD) in the rotating thin nanorod and nanocross, characterized by a splitting of $2\Omega$ in the resonance peaks. In contrast, the polarizability of the rotating disk with freely moving electrons in it do not show a resonance splitting. However, Eq.\ (\ref{disk}) predicts a difference in the decaying rate for different circular polarizations, which leads to a weak CD, manifested by the discrepancy in the linewidth and magnitude of the resonance peaks in Fig.\ \ref{Fig2}c.

Provided with the CD response of these rotating particles, we can readily conclude that linearly polarized illumination should exert optical torques on the particles, since the linearly polarized light can be decomposed into RCP and LCP components with equal amplitudes, which contribute with opposite and imbalanced torques. Figure \ref{Fig2}d,e,f shows rigorous results for the total optical torque exerted on the particles (Fig.\ \ref{Fig2}a,b,c) by linearly polarized light, calculated from Eq.\ (\ref{torque}) according to $M_{\rm dr}=M_+ + M_-$. In fact, for particles whose internal losses dominate over radiation losses, such as in Fig.\ \ref{Fig2}, Eq.\ (\ref{torque}) can be approximated as $M_\pm\approx 2\I{\alpha^\pm}|E_\pm|^2$, so that the torques in Fig.\ \ref{Fig2}d,e,f are given by $\I{\alpha^\pm}$, as shown in Fig.\ \ref{Fig2}a,b,c. For example, for rotating nanorod and nanocross under red-detuned laser illumination ($\omega<\omega_0$, Fig.\ \ref{Fig2}a,b), absorption of the LCP component should be strong compared with the RCP component ($\I{\alpha^-}>\I{\alpha^+}$), so that the total toque exerted by linearly polarized light should decelerate the particle rotation, leading to a RDC effect (blue shaded area in Fig.\ \ref{Fig2}d,e). Similarly, rotational acceleration (red shaded area in Fig.\ \ref{Fig2}d,e) is observed for blue-detuned illumination ($\omega>\omega_0$). The conditions for RDC and RDH in rotating nanorods and nanodisks are similar to those for their translational counterparts. This similarity originates in the fact that the polarizability observed in the rotating frame is equal to the motionless particle $\alpha^+(\omega_+)=\alpha^-(\omega_-)=\alpha(\omega,\Omega=0)$, so that the discussions for translational cooling based on Fig.\ \ref{Fig1}b are equivalently applied to these particles. However, for a rotating solid particle, such as the disk shown in Fig.\ \ref{Fig2}f, because of its different CD shown in Fig.\ \ref{Fig2}c, the RDH is observed for a nearly-resonant laser frequency $\omega$, while RDC is found at off-resonant laser frequencies in both blue- and red-detuned regimes.

\begin{figure*}[h]
\includegraphics[width=2.0\columnwidth]{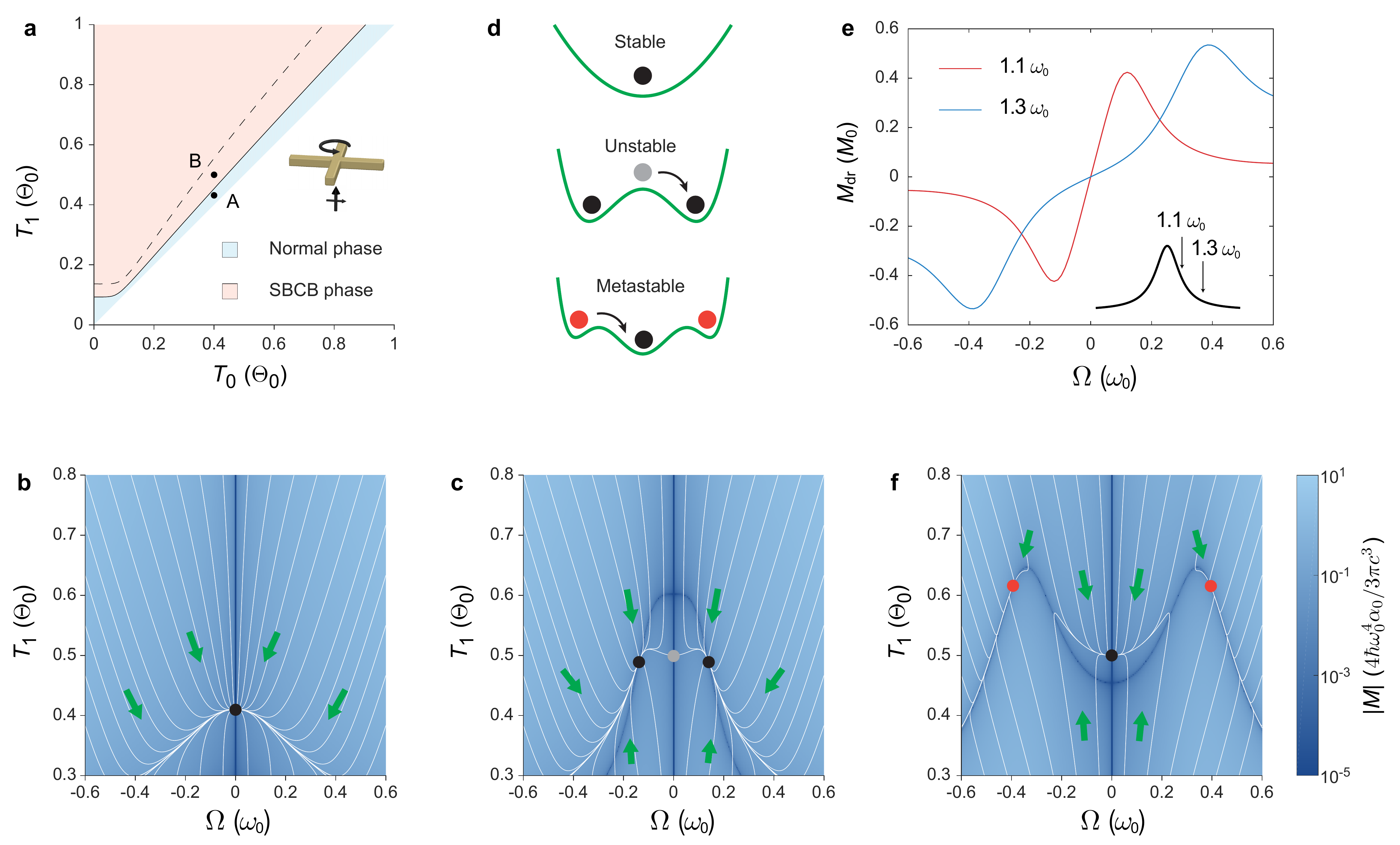}
\caption{\small {\bf Rotational dynamics of particles under linear polarized light.} {\bf a,} Stability of the nanocross considered in Fig.\ \ref{Fig2}b,e at rest under linearly polarized illumination of frequency $\omega=1.1\omega_0$ (solid curve) and $1.3\omega_0$ (dashed curve), with vacuum $T_0$ and particle $T_1$ temperatures normalized to $\Theta_0=\hbar\omega_0/k_{\rm B}$. For each $T_0$, a steady particle temperature $T_1$ is reached at a laser intensity $I(T_1)$. The black solid and dashed curves denote phase boundaries for frequency $\omega=1.1\omega_0$ and $1.3\omega_0$, respectively. {\bf b,c,} Evolution of the nanocross at arbitrary initial $T_1$ and $\Omega$ for light frequency $\omega=1.1\omega_0$, with vacuum temperature $T_0=0.4\Theta_0$ and laser intensities $I(0.41\Theta_0)$ ({\bf b}) or $I(0.5\Theta_0)$ ({\bf c}), which correspond to the black dots A and B in {\bf a}. {\bf d,} Illustration of the particle state in different phases. Top: the equilibrium state at $\Omega=0$ is stable (black dot, also in {\bf b}). Middle: the equilibrium state at $\Omega=0$ is unstable for higher light intensity (grey dot, also in {\bf c}). Bottom: a metastable configuration (see below). {\bf e,} Driving torque acting on the particle rotating at different velocities $\Omega$ for light frequencies $\omega=1.1\omega_0$ and $1.3\omega_0$. {\bf f,} Same with {\bf c} for light frequency $\omega=1.3\omega_0$. Two metastable configurations are observed, yielding the equilibrium at $\Omega=0$ stable (B point in {\bf a} is in normal phase for $\omega=1.3\omega_0$), which is intuitively illustrated in the lower panel of {\bf d}.
} \label{Fig3}
\end{figure*}
Besides the optical torque in Fig.\ \ref{Fig2}, the absorption process can increase the temperature of the particle, and the subsequent thermal emission produce a frictional torque on the particle if it is rotating \cite{Manjavacas2010}. This thermal frictional torque arises because the two circularly polarized dipoles $\pb_\pm$ posses different thermal populations and therefore there is an imbalance in the angular momentum released through thermal emission, leading to a frictional torque
\begin{align}
M_{\rm fr}=c\hbar\sum_{\nu=\pm}\nu \int  \dd\omega \rho^0(\omega)\sigma_{\rm abs}^\nu N_\nu(\omega)         ,\label{Mfr}
\end{align}
where $N_\pm(\omega)=n_1(\omega_\mp)-n_0(\omega)$ is the thermal imbalance of particle modes with vacuum, and $\rho_0(\omega)=\omega^2/3\pi^2 c^3$ is the projected local density of optical states in vacuum. For rotating particles, when the light frequency falls into the cooling regime, both thermal friction and optical cooling lead to the slowing down of rotation; provided the light frequency is in the heating regime, the driving torque $M_{\rm dr}$ exerted by the external illumination needs to exceed the thermal friction $M_{\rm fr}$ in order to produce acceleration.

An interesting phenomenon can be intuitively foreseen for a particle at rest under linearly polarized illumination when the light frequency is in the heating regime--- if the condition $M_{\rm dr}>M_{\rm fr}$ is satisfied near $\Omega=0$, a small particle rotation induced by any fluctuation can be amplified by RDH. Such effect implies the instability of the particle at rest, and considering the chiral symmetry of the system Hamiltonian, such instability manifests as a spontaneous chiral symmetry breaking (SCSB) process. To further analyze the particle stability at $\Omega=0$, we also need to find the steady state of the particle temperature under light irradiation, which is reached when the power absorbed from the laser, $P_{\rm abs}=\sum_\pm I_\pm\sigma_{\rm abs}^\pm$, is exactly compensated by the thermal-emission power. At arbitrary $\Omega$, the latter is given by \cite{Pan2019}
\begin{align}
P_{\rm ems}=c\hbar\sum_{\nu=\pm}\nu \int  \dd\omega \rho^0(\omega)\sigma_{\rm abs}^\nu N_\nu(\omega)           .\label{Pems}
\end{align}
The stability of the particle can be considered by taking the following aspects into account: (1) given the environment temperature $T_0$, the laser intensity is uniquely determined by the particle temperature $I(T_1)$; (2) the driving torque $M_{\rm dr}$ is then uniquely related to $T_1$ through $I(T_1)$; (3) $M_{\rm fr}$ depends on $T_0$ and $T_1$, and its magnitude compared with $M_{\rm dr}$ determines the stability. Following these considerations, at a given incident frequency $\omega$, the rotational stability of the particle can be mapped into $T_0$ vs $T_1$ plot.

In Fig.\ \ref{Fig3}a, we chose the nanocross in Fig.\ \ref{Fig2}b at rest ($\Omega=0$) as an example to illustrate its stability in the $(T_0,T_1)$ plane. This plot in fact constitutes a universal phase diagram, considering that the phase transition is featured by spontaneous symmetry breaking--- in the normal phase, the motionless state is stable; when increasing the intensity of illumination $I$ to heat the particle above the critical temperature dictated by the phase boundary in Fig.\ \ref{Fig3}a (black curves), the particle starts rotating spontaneously and the system enters into a SCSB phase.

To clearly reveal the features of the SCSB phase and find the final stable configration, we simulate the evolution of $\Omega$ and $T_1$ governed by the dynamical equations of motions $\dot{\Omega}=M_{\rm tot}/J$ and $\dot{T}_1=(P_{\rm abs}-P_{\rm ems}-M_{\rm tot}\Omega)/C$, where $M_{\rm tot}=M_{\rm dr}-M_{\rm fr}$, $J$ and $C$ are the moment of inertia and thermal capacity of the particle, respectively. The evolution of the system for vacuum temperature $T_0=0.4\Theta_0$ is shown in Fig. \ref{Fig3}b,c, taking the laser intensities $I(0.41\Theta_0)$ and $I(0.5\Theta_0)$ consistent with points A and B in Fig.\ \ref{Fig3}a.

For an incident frequency $\omega=1.1\omega_0$, the two points in Fig.\ \ref{Fig3}a fall into normal and SCSB phases, respectively (see boundary shown by black solid curve in Fig. \ref{Fig3}a). When the system is in the normal phase, as shown in Fig.\ \ref{Fig3}b, regardless of initial conditions of $\Omega$ and $T_1$, the system evolves toward a trivial stable equilibrium point (black dot) at $\Omega=0$ and $T_1=0.41\Theta_0$, since the laser intensity used is $I(0.41\Theta_0)$. By increasing the laser intensity to $I(0.5\Theta_0)$, the equilibrium point located at $\Omega=0$ is shifted to higher temperature at $T_1=0.5\Theta_0$ (grey dot). As expected, this equilibrium point becomes unstable, and the particle should start rotating toward a random direction, which eventually reaches one of the two new stable equilibrium states (black dots, Fig.\ \ref{Fig3}c). The characteristics of the normal and SCSB phases uncovered by these dynamics are intuitively illustrated in Fig. \ref{Fig3}d (upper and middle panels), where the black and grey dots correspond to the stable and unstable equilibrium states in Fig.\ \ref{Fig3}b and \ref{Fig3}c.

In Fig.\ \ref{Fig3}a, the phase boundary for $\omega=1.3\omega_0$ is also indicated through a dashed curve, which lies above the phase boundary for $\omega=1.1\omega_0$. As shown in Fig.\ \ref{Fig3}e, given a fixed laser intensity, the particle acquires larger optical torque from a laser at $\omega=1.1\omega_0$ at a small rotation $\Omega\approx 0$ compared to $\omega=1.3\omega_0$, so a laser with frequency $\omega=1.1\omega_0$ can more easily break the stability at $\Omega=0$. Point B for $\omega=1.3\omega_0$ lies within the normal phase, and the corresponding dynamical evolution is shown in Fig.\ \ref{Fig3}f. Although the equilibrium state at $\Omega=0$ in Fig.\ \ref{Fig3}f is stable, we also observe two metastable configurations (red dots) at relatively high rotation frequency, which arise because a larger driving torque can be exerted at a far-detuned frequency $\omega=1.3\omega_0$ for higher $\Omega$ (Fig.\ \ref{Fig3}e). In the metastable state, the particle can maintain its rotation, and only a large perturbation comparable to the energy barrier surrounding the metastable region can break the stability and force the particle to return to the equilibrium at $\Omega=0$, as illustrated in Fig.\ \ref{Fig3}d (lower panel).

We have generalized the Doppler cooling and heating mechanisms from translational to rotational degrees of freedom. We base our model on classical equations of motions and optical response, where oscillating dipoles can provide a general description for nanoparticles with various internal photonic or phononic excitations, as well as for molecules with electronic transitions. Specifically, a rotating nanorod can represent the optical response of diatomic or chain molecules, while the nanodisk is equivalent to molecules with two orthogonal dipole moments that can be freely coupled. To experimentally observe the phenomena here predicted, sparse particles or molecules trapped in high vacuum could be used to avoid gas friction and coupling between rotational and translational degrees of freedoms due to such scattering events. In experiment, concepts and techniques that are well-developed in current optical cooling setups, such as Zeeman splitting and chirping of the light frequency, could be combined with the mechanism here revealed to lead to actual applications. We also note that the optical response of a rotating nanoring is similar to a nanocross. Considering a ring particle with high electron mobility, such as a graphene nanoring, with its lattice fixed in the lab frame and exposed to linear polarized illumination, we expect a spontaneous electron current to arise, mimicking the mechanical rotation in our model. However, we argue that such spontaneous electron current cannot be achieved in a nanodisk, because CD as in Fig.\ \ref{Fig2}c cannot be observed in such case. Our work unveils fundamental mechanisms enabling novel optical trapping techniques, and also offers new insights into the optical response of out-of-equilibrium rotational systems.

\noindent
\\
\textbf{Acknowledgements}
\noindent This work has been supported in part by the Spanish MINECO (Grants No. MAT2017-88492-R and No. SEV2015-0522), ERC (Advanced Grant No. 789104-eNANO), the Catalan CERCA Program, and Fundaci\'{o} Privada Cellex.


\end{document}